\documentclass{jps-cp}
\usepackage{txfonts} %Please comment out this line unless the txfonts package is availabe in your LaTeX system.
\usepackage{graphicx}
\usepackage{color}
\usepackage{footmisc}

\title{Monero Peer-to-peer Network Topology Analysis}

\author{Yu \textsc{Gao}$^{1}$, Yu \textsc{Zhang}$^{1}$, Matija \textsc{Pi\v{s}korec}$^{1}$ and
Claudio \textsc{J. Tessone}$^{1}$}

\inst{$^{1}$Blockchain Distributed Ledger Technologies, IfI, University of Zurich, Zurich, Switzerland}

\email{yugao@ifi.uzh.ch}

\recdate{March 4, 2025}

\abst{ 

Monero, a privacy-focused cryptocurrency, employs a decentralized peer-to-peer (P2P) network that plays a critical role in transaction propagation and consensus formation. While much research has explored Monero’s privacy transaction mechanisms, its underlying P2P network architecture has remained relatively underexplored. In this study, building on our recent work on Monero network detection, we further investigate the network topology of Monero's P2P structure, which has evolved following recent protocol updates that enhanced security by obscuring peer information. Using $k$-core decomposition, we confirm that the Monero network exhibits a core-periphery structure, where a tightly interconnected core of supernodes is crucial for maintaining network cohesion, while peripheral nodes rely on these core nodes for connectivity. This structure explains why targeting central nodes does not easily lead to the rapid disintegration of the network's largest connected component while also providing a deeper understanding of the true architecture of Monero's peer protocol.}

\kword{Monero P2P Network, Network Topology, Core-Periphery}

\begin{document}
\maketitle

\section{Introduction}

The concept of decentralized transaction ledger was first introduced in a seminal 2008 work by Satoshi Nakamoto~\cite{nakamoto2008bitcoin}, laying the foundation for trustless digital currencies. In Nakamoto’s system, there is no central authority responsible for securing and maintaining the ledger. Instead, it is collectively maintained by all network participants, with each node storing a complete copy and appending new transactions to an immutable and ever-growing log. This structure, known as a blockchain, organizes transactions into blocks, which are cryptographically linked to ensure integrity. The introduction of Bitcoin demonstrated that a decentralized digital asset could function as a store of value and a medium of exchange, fundamentally transforming our approach to digital trust\cite{buterin2014, schutze2006}.

A critical element of Distributed Ledger Technology (DLT) is the peer-to-peer (P2P) network, which serves as the backbone for transaction verification and propagation\cite{biryukov2014, decker2013}. Without a P2P structure, the concept of decentralization would not be possible. The decentralized nature of the P2P network eliminates central intermediaries, distributing control and responsibility across all participants\cite{buford2009, bach2018}. This ensures that the ledger remains resilient, transparent, and secure, with each participant playing an essential role in maintaining its integrity\cite{heilman2015, neudecker2015, gao2024 
}.

Bitcoin’s success catalyzed the development of a broader ecosystem of digital currencies. Among these blockchain-based cryptocurrencies, Monero stands out for its strong emphasis on privacy and anonymity. Unlike Bitcoin, where transactions recorded on a public ledger are traceable which facilitates user transaction analysis \cite{sornette2025transaction,zhang2024bitcoin}, Monero employs advanced cryptographic techniques such as ring signatures, stealth addresses, and Confidential Transactions to obfuscate sender, receiver, and transaction amounts\cite{moser2017, miller2017, eskandari2018}. While extensive research has been conducted on Monero’s privacy mechanisms, its peer-to-peer (P2P) network architecture, which plays a crucial role in transaction propagation and consensus formation, remains largely unexplored. Given that the structure and resilience of a P2P network directly influences security, decentralization, and resistance to attacks, understanding Monero’s network topology is essential.

% When a new node joins the Monero network, it initially connects to predefined seed nodes to discover and integrate with other active nodes. The node then employs a handshake protocol to validate connections and synchronize data, including blocks and peer lists. Once connected, the node maintains and updates its peer list, managing incoming connections to ensure network stability. Each Monero peer maintains two lists: \textit{white\_list} and \textit{gray\_list}. The \textit{white\_list} holds up to 1000 recent handshake peer addresses, while the \textit{gray\_list} contains up to 5000 addresses of peers that did not respond or have older timestamps.

Our findings in recent work revealed that the largest connected component of the Monero P2P network is highly resilient to targeted attacks on central nodes, as supernodes exhibit strong connectivity. Furthermore, we also applied community detection, a key technique in network science, but it did not reveal any clear community divisions. 

These observations may suggest the presence of a core-periphery structure, where a tightly connected core of supernodes plays a pivotal role in maintaining network cohesion, while peripheral nodes rely on these core nodes for connectivity\cite{borgatti2000, holme2005core}. The $k$-core decomposition provides valuable insights into network topology and has been applied in various domains, including social network analysis, biological systems, and communication networks. The hierarchical structure of $k$-cores allows for distinguishing between peripheral nodes with low connectivity and central nodes that contribute to the structural integrity of the network.

In this study, we systematically investigate whether a core-periphery structure exists in the Monero P2P network and analyze its properties if a core-periphery structure exists. For detecting the core-periphery structure, we mainly employ the $k$-core algorithm from Network Science\cite{barabasi2013, newman2018} to explore and quantify the network’s structural organization. Furthermore, we delve into the internal network topology, examining how core nodes connect with peripheral nodes and the connectivity patterns within the core community.

Our analysis confirms our hypothesis about the Monero network structure, which follows a core-periphery pattern. This finding deepens our understanding of Monero’s underlying P2P architecture and contributes to both theoretical insights and practical security considerations for auto-peering protocols in blockchain technologies.

\section{Monero Network Architecture Detection Methods}

Before 2019, Monero's protocol required each peer to share a list of the top 250 peers in their \textit{white\_list} sorted by the most recent timestamp, including timestamp data. Following Cao et al.'s findings~\cite{cao2020exploring}, the Monero protocol was updated to improve security: the returned list of peers now comprises 250 peers randomly selected among the top 300 addresses in the \textit{white\_list}, ranked by the most recent timestamp, but no timestamp information is returned\footnote{https://www.getmonero.org}.
 This change obscures network details, making previous network inference methods introduced in \cite{cao2020exploring} ineffective. In our prior work\cite{charted25}, we mapped the Monero P2P network with relatively high accuracy using the classic $k$-means method, and then analyzed its connectivity patterns through centrality metrics and visualization techniques. However, this is just a partial picture of Monero's P2P network, which is not enough to understand  Monero’s network topology and its implications for decentralization, security, and potential vulnerabilities. In the following parts, we explore its network architecture further.

\subsection{Nearest Neighbour Degree}

In our previous work, some properties of Monero's P2P network have been investigated, including the degree distribution (degree's cumulative density distribution) and the relationship between peers' degrees and betweennesses. As a complement, it is crucial to examine connection properties, particularly degree correlation with nodes' nearest neighbor degree ($K_{nn}(k)$)\cite{barabasi2013}.

Consider each Monero peer as a node in the graph \( G \). If peer \( i \) and peer \( j \) are neighbors, then there is an edge connecting \( i \) and \( j \) in \( G \). We use $\mathbf{A} =(A_{ij})$ to represent the adjacency matrix of $G$, where $A_{ij} = 1$ if nodes $i$ and $j$ are neighbors and $A_{ij} = 0$ otherwise. Then, $K_{nn}(k)$ is defined based on following equations:

\begin{align}
    K_{nn}(k) &= \frac{k_{nn}(k_i)\cdot \delta(k_i=k)}{N_k},\\
    k_{nn}(k_i) &= \frac{\sum_{j=1}^{N}A_{ij}\cdot k_j}{k_i},
\end{align}
where $A_{ij}$ denotes the entry of $i^{th}$ row $j^{th}$ column in the adjacency matrix; $k_i$ denotes the degree of $i^th$ node (peer in this paper); $k_{nn}(k_i)$ indicates the average degree of $i^th$ node's all neighbours; $\delta(k_i=k)$ is a Dirichlet function which is 1 if $k_i=k$ and 0 otherwise; $N_k$ is the number of nodes whose degree is $k$. 
$K_{nn}(k)$ reveals how peers of a given degree connect to others across the network and the $K_{nn}(k)$ of Monero's P2P network is shown in Fig. \ref{knnk}.

\subsection{$k$-core Alogrithm}

In network science, the \textit{core} of a network refers to a densely connected substructure that plays a critical role in maintaining the network’s connectivity and function\cite{newman2018}. One widely used method for identifying such a core is $k$-core decomposition, which reveals hierarchical layers of node connectivity. We employ the $k$-core algorithm\cite{batagelj2003} to detect discrete versions of core-periphery structure in the Monero network. 
% Fig.\ref{core-peri} is the Matrix plot of the whole Monero network we inferred.

A $k$-core of a graph \( G = (V, E) \) is defined as a maximal subgraph in which every node has at least \( k \) connections (degree \(\geq k\)) within that subgraph. The $k$-core structure is obtained through an iterative process, where nodes with degrees less than \( k \) are progressively removed until all remaining nodes satisfy the degree constraint. The \textit{core number} of a node is the largest \( k \) for which it belongs to a $k$-core.

% \section{Core-Periphery Detection}

\section{Monero Network Structure Detection Result}

\begin{figure}[htbp]
\includegraphics[width=0.8\textwidth]{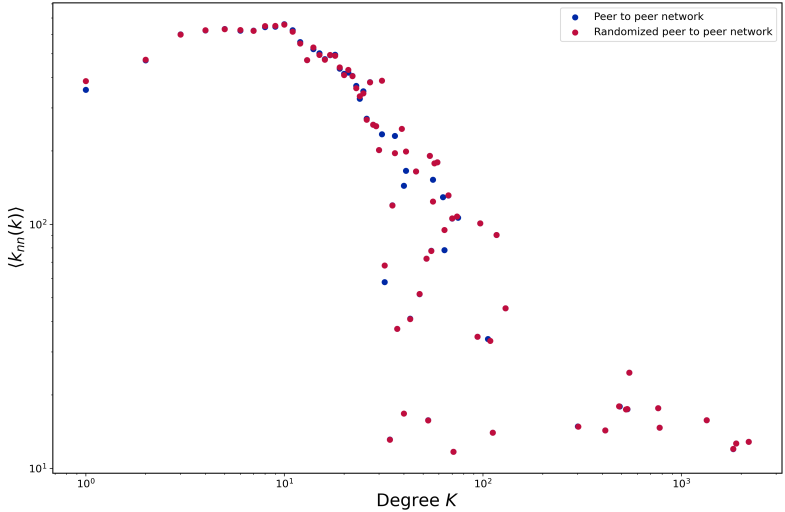}
\caption{$K_{nn}(k)$ of Monero's P2P network with assortativity -0.28.}
\label{knnk}
\end{figure}

Fig.\ref{knnk} shows the relationship between peers' degrees and their nearest neighbor degrees ($K_{nn}(K)$) in Monero's P2P network.
For $K<10$ where $K$ denotes peers' degree, $K_{nn}(K)$, starting from an extremely large value of approximately 300, exhibits an upward trend and hits 600 when $K=10$, which indicates that peers with degree less than 10 are more likely to connect to other peers with extremely large degrees, such as some super nodes. For $K>10$, $K_{nn}(K)$ decreases with peers' degrees, which again indicates that peers with smaller degrees tend to connect to those peers with large degrees. The assortativity value is -0.28, which is consistent with our analysis of $K_{nn}(K)$. The disassortativity of Monero's P2P network may partially reflect its core-periphery structure which we will investigate in detail below.

As inferred from the paper mentioned above, the Monero network exhibits a distinct core-periphery structure, with 4,837 nodes in total. Of these, 3,153 neighbors are directly connected to the 14 super nodes, which have the largest degree of centrality. The 14 super nodes and their direct neighbors' mutual connections represent 82.1\% of the entire network, highlighting their central role in the core of the P2P network.

The analysis of the P2P network in the last paper reveals a highly interconnected core, where top-degree nodes share over 91\% of their direct neighbors, with 9 out of 14 reaching nearly 100\%. This indicates strong network resilience. Peripheral nodes act as endpoints with minimal connections, while core nodes—mainly supernodes or public seed nodes—facilitate efficient data routing.

%
% \subsection{K-core alogrithm implementation}

\begin{figure}[htbp]
\includegraphics[width=0.9\textwidth]{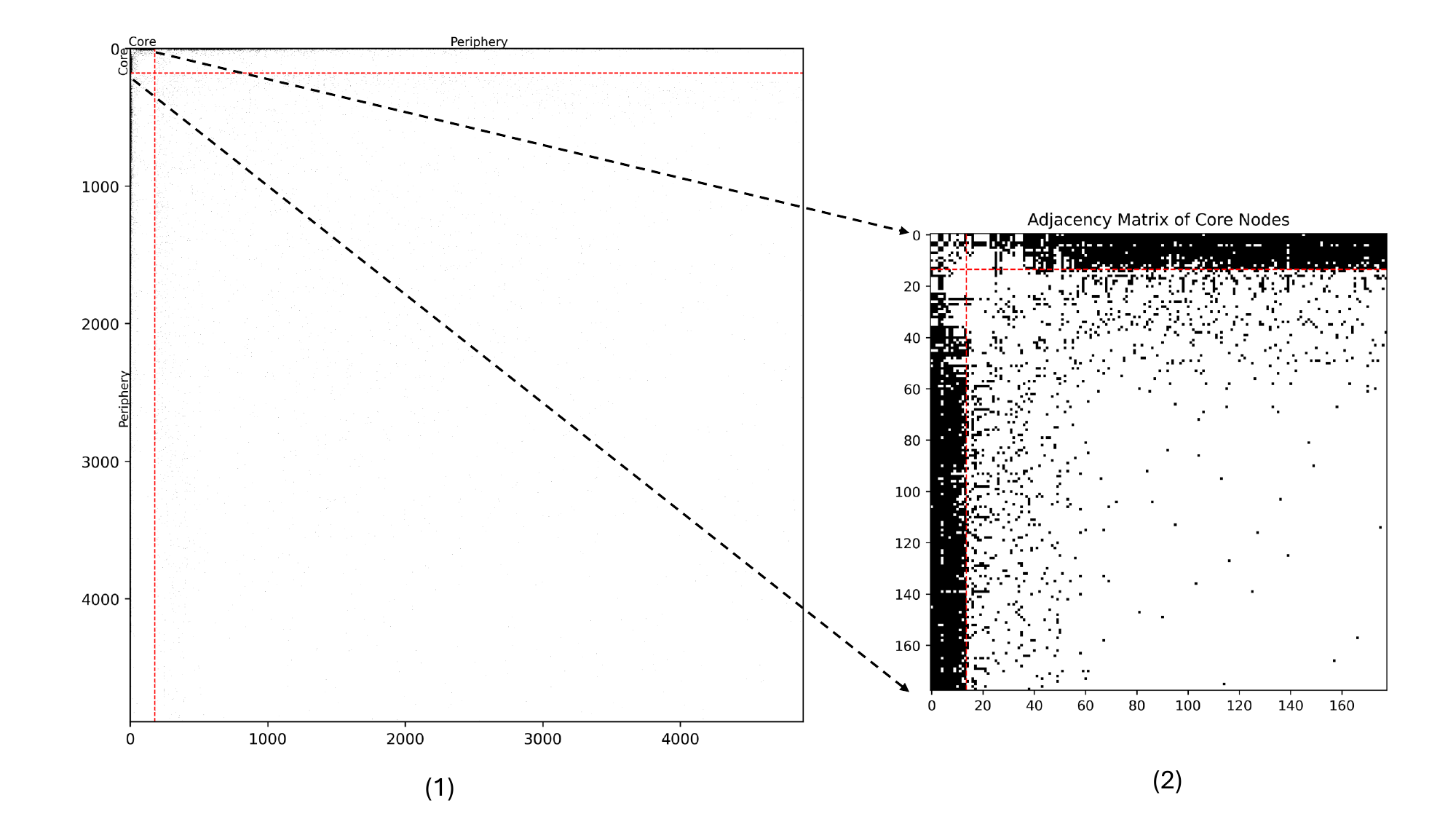}
\caption{(1) Core-periphery structure of the Monero Peer-to-Peer network using the $k$-core algorithm. Nodes are reordered based on their degree. Filled cells indicate the presence of edges, while open cells denote their absence. The dotted lines mark the boundary between the core and periphery. (2) The zoomed-in Matrix plot of the core part of the Monero P2P network.}
\label{core-peri}
\end{figure}

The $k$-core algorithm partitions the Monero network into a core and a periphery, as shown in Fig. \ref{core-peri}. The detected structure suggests a core-periphery organization, where core nodes are densely interconnected, while peripheral nodes have fewer connections. The matrix representation illustrates this structure, with nodes reordered by degree. Filled cells indicate the presence of edges, while open cells denote their absence. The dotted lines mark the boundary between the core and periphery. 

% {\color{red}{need to tell the procedures to draw figure \ref{core-peri}, like nodes' sorting orders}}

By detailed checking, we find that
the $k$-core with $k = 16$ consists of 178 peers, forming the most interconnected subgraph within the network. An analysis of its adjacency matrix reveals distinct structural properties, as shown in Fig. \ref{core-peri} (2).
A notable observation is that the 14 super-peers with the highest degree predominantly establish connections with non-super-peers rather than with each other. This is evident from the relatively sparse connectivity among super-peers in the top-left region of Fig. \ref{core-peri} (2), compared to the denser connections observed in the lower and right-hand regions. This suggests a hierarchical topology where super-peers function as relay nodes, primarily facilitating communication between non-super-peers rather than forming a tightly connected clique.

\begin{figure}[htbp]
\includegraphics[width=0.7\textwidth]{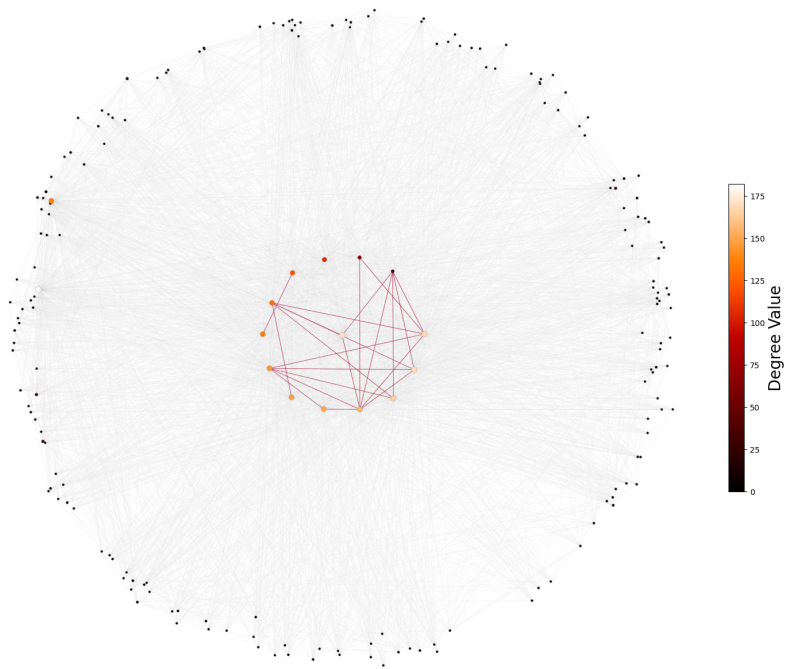}
\caption{The graph of the core part corresponding to Fig.\ref{core-peri} (2). These 14 super-peer nodes with larger sizes are located in the center. The color of a node varies with the magnitude of its degree, using the 'gist\_heat' colormap. The smaller the degree, the darker the color.}
\label{network-visual}
\end{figure}

To further illustrate this structure, the $k$-core subgraph is visualized in Fig. \ref{network-visual}. Here, the 14 super-peers, represented as larger nodes, act as hubs, each directly linked to multiple non-super-peers. However, their interconnectivity remains limited.

Furthermore, when these 14 super-peers are removed from the $k$-core, the remaining peers exhibit significantly reduced connectivity, with many becoming isolated or only weakly connected. This indicates that the core's structural robustness is highly dependent on these super-peers, reinforcing their critical role in maintaining overall network connectivity. Such a topology is characteristic of unstructured P2P networks, where a small set of high-degree nodes serves as the backbone for efficient data propagation and peer coordination.

% \begin{figure}[htbp]
% \includegraphics[width=0.5\textwidth]{fig/adjacency_matrix_kcore_core_plot.png}
% \caption{The zoomed-in Matrix plot of the core section in Fig.\ref{core-peri} of the Monero P2P network.}
% \label{core-core-peri}
% \end{figure}

\begin{figure}[htbp]
\includegraphics[width=0.7\textwidth]{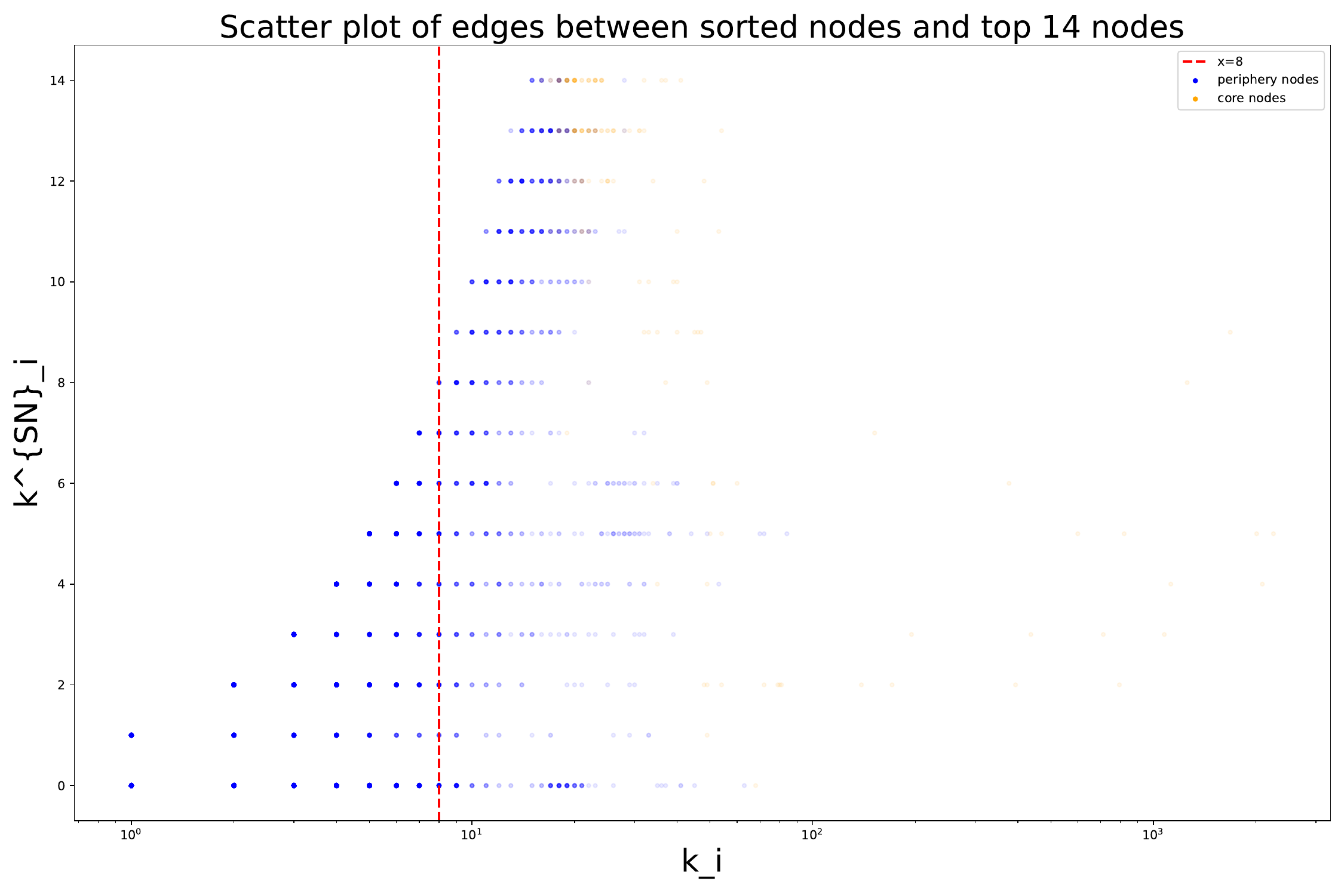}
\caption{Scatter plot of node connectivity, where the x-axis ($k_{i}$) represents the total node degree on a logarithmic scale and the y-axis ($k\{SN\}_{i}$) indicates the number of connections each node has to the top 14 high-degree nodes. Green points denote nodes that belong to the network's k-core, while blue points represent nodes outside the k-core. Dashed red and purple lines mark the reference thresholds at $x = 8$.}
\label{scatter}
\end{figure}

The scatter plot in Fig.\ref{scatter} visualizes the relationship between each node’s total degree $k_{i}$ and the number of connections it has to the top 14 highest-degree nodes $k\{SN\}_{i}$. The x-axis represents the total degree of nodes on a logarithmic scale, while the y-axis represents the number of edges each node has with the top 14 nodes. Nodes are color-coded based on whether they belong to the k-core subgraph—green indicates k-core nodes, while blue represents non-k-core nodes. Additionally, two dashed reference lines are plotted at $x=8$.

 The presence of a dense cluster of nodes around $x=8$ suggests that a significant portion of the network has exactly or close to 8 connections. This pattern likely arises from a default network configuration where nodes, if not explicitly configured otherwise, establish 8 outgoing connections. Such behavior is common in peer-to-peer (P2P) networks, where nodes follow a pre-defined number of connections for efficiency and stability.

A shallow distribution of core nodes (if most core nodes are bunched together in the same region) may imply that once nodes enter the core, they share a similar connectivity pattern. This can be good for internal robustness, but also indicates that the network might rely on a fairly uniform set of connection rules or behaviors.

% \vspace{-1cm}
\section{Conclusion and Discussion}

Our analysis of Monero’s P2P network architecture reveals a distinct core-periphery structure, where a small set of super-peers plays a central role in maintaining network connectivity and facilitating data propagation. By examining the degree correlation, we observe a disassortative topology, where low-degree peers preferentially connect to high-degree super-peers. The network’s negative assortativity coefficient further supports this, indicating that Monero's topology resembles a hierarchical structure rather than a purely decentralized one.

With $k$-core algorithm, we detected a core-periphery structure. 
These super-peers exhibit limited interconnectivity but serve as major hubs for peripheral nodes. Their removal drastically reduces the connectivity of the remaining peers, highlighting their structural significance in the network. This suggests that while Monero’s P2P network remains functionally decentralized, its resilience is highly dependent on a small subset of influential nodes.

Overall, our findings provide valuable insights into the topological vulnerabilities of Monero’s network. The reliance on super-peers may introduce potential attack vectors, such as targeted disruptions aimed at these critical nodes. Future work could focus on measuring network robustness under adversarial conditions and exploring potential mechanisms to enhance decentralization and resilience.

\end{document}